\documentclass[11pt]{article} 
\usepackage{amsmath}
\usepackage{times}
\usepackage{subfigure}
\usepackage{epsfig}
\usepackage[english]{babel}
\usepackage{graphicx}
\usepackage{hyperref}
\usepackage{color}
 
\newcommand{\amp}{WA~}

\title{Nonlinear dynamics of running: Speed, stability, symmetry and the effects
   of leg amputations}

\author{Nicole Look\\University of Colorado, Boulder \\ Department of
  Applied Mathematics \and Christopher J. Arellano \\University of
  Colorado, Boulder \\ Department of Integrative Physiology \and Alena M.
  Grabowski\\University of Colorado, Boulder \\ Department of
  Integrative Physiology \and William J. McDermott\\The Orthopedic
  Specialty Hospital, Murray, UT, USA \and Rodger Kram\\University of Colorado,
  Boulder \\ Department of Integrative Physiology \and Elizabeth
  Bradley\\University of Colorado, Boulder \\ Department of Computer
  Science \\
\\
In review, {\sl CHAOS}.}
  
\date{}
\begin{document}
 
\maketitle
 
\begin{abstract}

In this paper, we study dynamic stability during running, focusing on
the effects of speed and the use of a leg prosthesis.  We compute and
compare the maximal Lyapunov exponents of kinematic time-series data
from subjects with and without unilateral transtibial amputations
running at a wide range of speeds.  We find that the dynamics of the
affected leg with the running-specific prosthesis are less stable than
the dynamics of the unaffected leg, and also less stable than the
biological legs of the non-amputee runners.  Surprisingly, we find
that the center-of-mass dynamics of runners with two intact biological
legs are slightly less stable than those of runners with amputations.
Our results suggest that while leg asymmetries may be associated with
instability, runners may compensate for this effect by increased
control of their center-of-mass dynamics.

\end{abstract}
 
\clearpage

{\noindent \bf Lead Paragraph}

\bigskip

In order to understand the combined effects of speed, stability, and
the use of leg prostheses, it is important to explore the dynamical
details of running.  Nonlinear time-series analysis of kinematic gait
data can effectively elucidate these details.  There have been a
number of experimental studies of the dynamics of running (e.g.,
\cite{jordan}).  To our knowledge, however, no one has explored the
stability dynamics of runners with leg amputations, a population to
whom dynamical stability seems an especially important issue.  Using
nonlinear time-series analysis on motion-capture data from treadmill
studies, we analyzed the gait dynamics of runners with and without a
unilateral transtibial amputation, from a slow run up to each
individual's top speed.  We used standard delay-coordinate embedding
techniques to reconstruct the dynamics from scalar time-series traces
of the positions of various anatomical markers (e.g., the height of
the sacrum or the sagittal-plane angle of the right knee), then we
calculated the maximal Lyapunov exponent $\lambda_1$ of each resulting
trajectory.  We found that stability decreased at faster speeds for
all runners, with or without amputations.  We also found that
lower-limb dynamics were less stable (viz., higher $\lambda_1$) for
the affected leg of runners with an amputation than for their
unaffected leg---and less stable than {\sl either} leg of the
non-amputee runners.  The $\lambda_1$ values increased with running
speed, but the inter-leg and inter-group relationships remained
largely the same.  Surprisingly, the results showed that the
center-of-mass dynamics of non-amputee runners were slightly less
stable than for runners with a unilateral transtibial amputation.
This suggests that asymmetries may lead to
instability in the leg dynamics that are compensated for by increased
control of the center of mass.

\section{Introduction}

Analysis of the dynamics of locomotion elucidates temporal variations
in gait patterns and also leads to a better understanding of
stability.  Nonlinear time-series analysis techniques have been used
to study various aspects of human walking, including differences
between normal and pathological walking gait (e.g.,
\cite{dingwell00,hausdorff09}), the effects of age and illness
\cite{buzzi03,scafetta09}, synchronization when two people walk
side-by-side \cite{nessler09}, recognition of an individual from his
or her gait \cite{frank10}, and stability of walking in the face of
continuous perturbations \cite{dingwell11}.  The goal of our study was
to explore the effects of speed, stability, and leg prosthesis use in
the dynamics of a different locomotion pattern: running.  At moderate
speeds, a runner can be modelled as a bouncing spring-mass system,
whereas walking can be represented as a series of inverted-pendulum
arcs.  A number of interesting models of the dynamics of running have
been developed in the biomechanics, robotics, and nonlinear dynamics
communities (e.g., \cite{holmes06}), some of which were specifically
constructed to explore stability issues \cite{carver09}.  Only a few
studies involved nonlinear analysis of laboratory data from human
runners (e.g., \cite{mcgregor09}), but none have explored the temporal
details of the dynamics.  Further, the effects of prosthesis use on
these dynamics have not, to our knowledge, been studied at all.
%
%
%

To explore these dynamics, we collected data from 17 subjects running
on an instrumented treadmill across a wide range of speeds (3-9 m/s).
Six of these subjects had a unilateral transtibial amputation and
eleven had two intact biological legs.  The time-series data included
the $xyz$ positions of reflective markers placed on the body, gathered
via motion-capture cameras over a number of gait cycles.  We
reconstructed the center-of-mass dynamics using delay-coordinate
embedding on various scalar projections of these raw data.  We
reconstructed the limb dynamics by converting the 3D positions to
joint angles and then embedded those angle traces.  Finally, we
calculated the maximal Lyapunov exponent $\lambda_1$ of each of the
embedded trajectories using the algorithm of Kantz \cite{kantz94}.
Quantifying the dynamic stability of human locomotion, which is
defined as resistance to change under perturbation, is not trivial.
Full {\sl et al.}, for instance, suggested a detailed approach that
decomposes each trajectory into limit cycles and quantifies the rates
of recovery from perturbations in different state-space directions
\cite{full02}.  For the purposes of a time-series analysis study with
finite amounts of data, $\lambda_1$ is a proxy for stability that has
been used extensively in human walking studies (e.g.,
\cite{bruijn2009a,bruijn2009b,dingwell00,lockhart}).


The approach outlined in this paper is useful not only for exploring
the nonlinear dynamics of running, but also for assessing the
sensitivity of those dynamics to perturbations.  A better
understanding of these effects could inform the design of better
prostheses for this activity.  A careful assessment of dynamics is
also useful for understanding the intertwined roles of symmetry and
stability.  Seeley {\sl et al.} \cite{seeley08} and Gundersen {\sl et
  al.} \cite{gundersen89}, for instance, demonstrated that healthy
walking gait is bilaterally symmetrical, even though slight
asymmetries may develop to accommodate for changing environmental
factors.  Skinner \& Effeney \cite{skinner85} found significant
bilateral asymmetries in the lower-limb kinematics of people with leg
amputations during walking; Enoka {\sl et al.}  \cite{enoka} found
similar asymmetries in running.  During running and sprinting,
Grabowski {\sl et al.} determined that people with a unilateral
transtibial amputation applied significantly less force to the ground
with their affected leg than their unaffected leg \cite{grabowski10}.
It is not known, however, if that kind of force asymmetry affects the
dynamic stability of gait.  Variability and asymmetry are not
necessarily detrimental; in the introduction to the 2009 focus issue
of {\sl CHAOS} on ``Bipedal Locomotion—--From Robots to Humans,''
Milton \cite{milton09} writes, ``Thus it is possible that a certain
amount of kinematic variability in certain aspects of performance
might be indicative of a healthier dynamical system.''  A comparison
of the gait dynamics of non-amputee runners to those of runners with a
unilateral transtibial amputation may elucidate these subtle effects.

\smallskip
\noindent The research reported in this paper was driven by three hypotheses:
\label{page:hypotheses}
\begin{enumerate}
\item For individuals with or without a transtibial amputation, dynamic
  stability will decrease at faster running speeds.

This hypothesis is based on the work of England \& Granata, who found
that faster walking speeds lead to larger $\lambda_1$ (viz., less
stability) \cite{england07}.  We expected a similar relationship
between speed and stability during running.

\item The $\lambda_1$ of the lower-limb dynamics of runners with a
  unilateral transtibial amputation will be asymmetric, across all
  speeds.

This followed from the geometric asymmetry of the dynamical system,
defined as the notable anthropomorphic differences (mass and moment of
inertia) between the affected and unaffected legs, as well as the loss
of muscular control in the affected leg.

\item The $\lambda_1$ of the center-of-mass dynamics of runners with a
  unilateral transtibial amputation will be greater than in
  non-amputee runners.

We based this hypothesis on the rationale that symmetry in the lower
limbs poses a challenge to maintaining overall stability during
locomotion.

\end{enumerate}

\noindent Our study confirmed our first two hypotheses.  The
$\lambda_1$ values increased with running speed, while the inter-leg
and inter-subject relationships remained largely the same across
speed.  We found that lower-limb mechanics were generally less stable
(viz., higher $\lambda_1$) for the affected leg of runners with
amputations than for their unaffected leg---or than for {\sl either}
leg of the non-amputee runners.  Surprisingly, though, our results
showed that the center-of-mass dynamics of non-amputee runners were
slightly {\sl less} stable than in runners with a unilateral transtibial
amputation.

The following two sections describe how the data for this study were
collected and analyzed.  The results are presented in
Section~\ref{sec:results} and discussed in
Section~\ref{sec:discussion}.  

\section{Data Collection}

\begin{figure}
   \centering
     \includegraphics[width=0.8\textwidth]{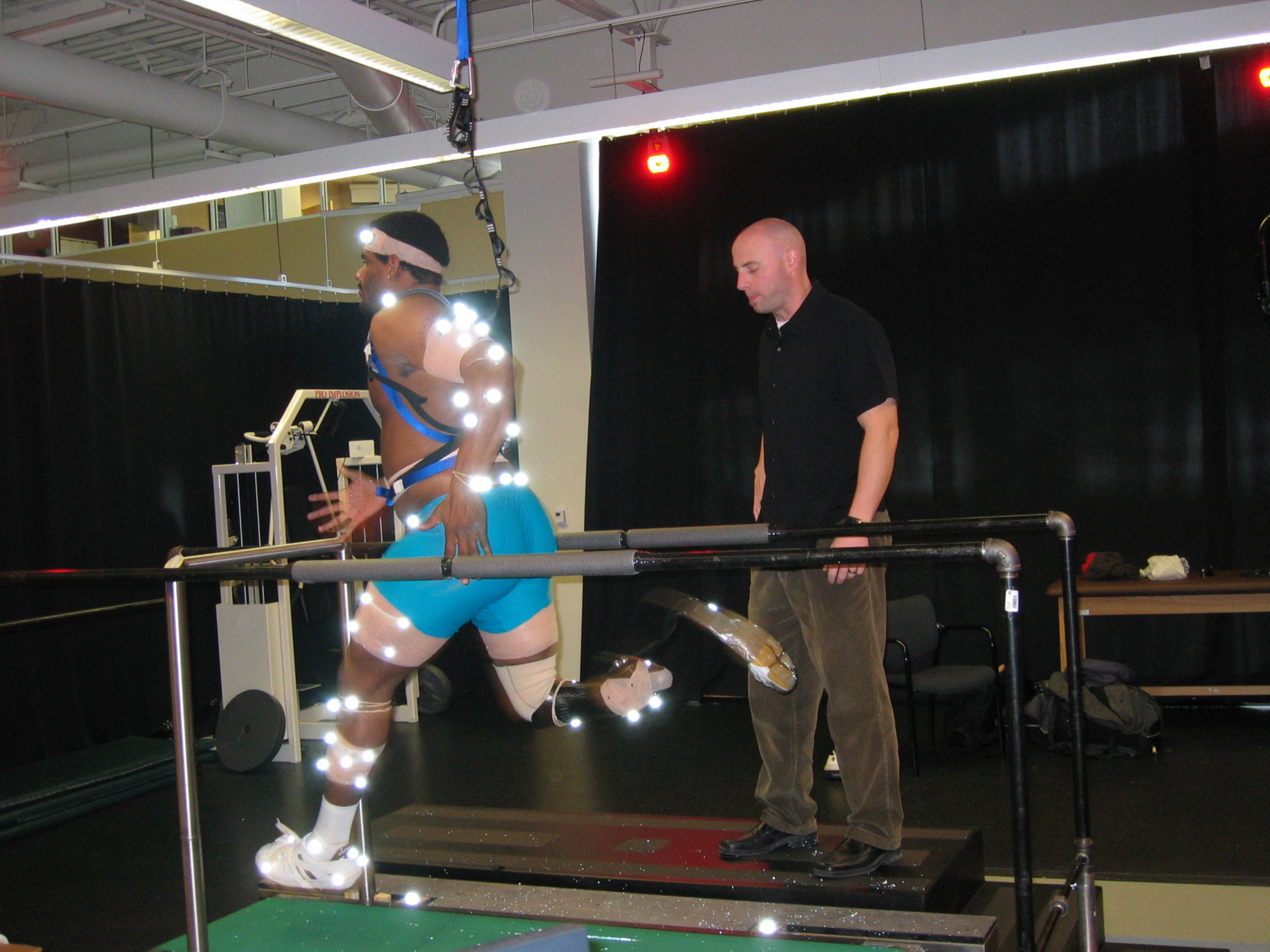}
     \caption{Subject with a unilateral transtibial amputation running
       on a high-speed instrumented treadmill}
 \label{fig:setup}
 \end{figure}

A total of 17 subjects---6 runners (4 male and 2 female) with a
unilateral transtibial amputation and 11 runners (8 male and 3 female)
without amputations---participat\-ed in the study described in this
paper.  In the rest of this document, members of these two groups are
designated with the acronyms \amp and NA, respectively.  All of the
experiments occurred at the Biomechanics Laboratory of the Orthopedic
Specialty Hospital (Murray, Utah).  A photograph of the setup is shown
in Figure~\ref{fig:setup}.
All subjects gave informed written consent according to the
Intermountain Healthcare IRB approved protocol.  Each \amp subject
used his or her own sprint-specific passive-elastic prosthesis during
the tests.  We measured each subject's height, mass, prosthesis mass,
and standing leg lengths.  We defined leg length as the vertical
distance from the greater trochanter to the floor during standing.  We
measured the length of the affected leg of each \amp subject when it
was unloaded, by having the individual stand with a 2'' wooden block under
the unaffected leg.

Subjects performed a series of constant-speed running trials on a
custom high-speed treadmill (Treadmetrix, Park City UT).  Each trial
consisted of at least ten strides except for top-speed trials, which
consisted of 8 strides.  After a brief warm up, subjects started the
series of running trials at 3 m/s.  Each subsequent trial speed was
incremented by 1 m/s until subjects reported that they were
approaching their top speed.  Smaller speed increments were then
employed until subjects reached their top speed, defined as the speed
at which they could not maintain their position on the treadmill for
more than eight strides \cite{Weyand}.  Subjects were allowed as much
time as desired between trials for full recovery.  The pelvis position
was defined by reflective markers attached to the anterior and
posterior iliac spines and iliac crests of the right and left sides.
We used reflective marker clusters to define the thigh and shank
segments.  In order to define the hip and knee joint centers, we also
placed reflective markers on the greater trochanters and the medial
and lateral femoral condyles of the right and left legs.  We used a
marker placed over the sacrum as a proxy for the center of mass.  We
used motion-capture cameras (Motion Analysis Corporation, Santa Rosa,
CA) to measure the 3D positions of those markers at a rate of 300
frames per second, then calculated the joint angles from those data
using Visual3D software (C-Motion Inc., Germantown, MD).
%
%
We did not normalize the timebase of each data set to the average
stride period, as is done in some gait studies, 
because that operation would obscure the speed effects in which we were
interested.

\label{page:COM-sacrum} We used the spatial position of
the sacrum marker at the base of the spine to study the center-of-mass
dynamics.  This is, of course, an approximation.  The sacrum location
is close to the overall body center of mass (COM) when a person stands
in the standard anatomical position.  However, when a person runs,
their COM position moves within the body.  It is possible to estimate
the COM location using a segmental approach, but this methodology
relies on many assumptions and estimates about human body segment
parameters.  Furthermore, there are no established methodologies for
estimating how an amputation and/or the use of a running-specific
prosthesis affects the position or movement of the COM.  Thus, we
chose to use the sacrum marker as a proxy for estimating the COM
location and studying its dynamics during running.


\section{Time-Series Analysis}
\label{sec:tsa}

The time-series data described in the previous section comprised
time-series traces of dozens of joint positions in 3D space.  To
reconstruct the locomotion dynamics from these data, we used
delay-coordinate embedding.  Provided that the underlying dynamics and
the observation function $h$ that produces the measurement $x(t)$ from
the underlying state variables $X$ of the dynamical system are both
smooth and generic, the delay-coordinate map
\begin{equation}\label{eqn:takens}
F(\tau,m)(x) = ([x(t) ~ x(t+\tau) ~ \dots ~x(t+(m)\tau)])
\end{equation} 
with delay $\tau$ from a $d$-dimensional smooth compact manifold $M$
to $R^{m}$ is a diffeomorphism on $M$ if the embedding dimension $m$
is greater than $2d$ \cite{takens}.  Here, $M$ is the dynamics of the
human body; $h$ is the measurement executed by the motion-capture
system, plus the post-processing involved in the conversion from
marker positions to joint angles.

\begin{figure}
   \centering
     \includegraphics[width=0.8\textwidth]{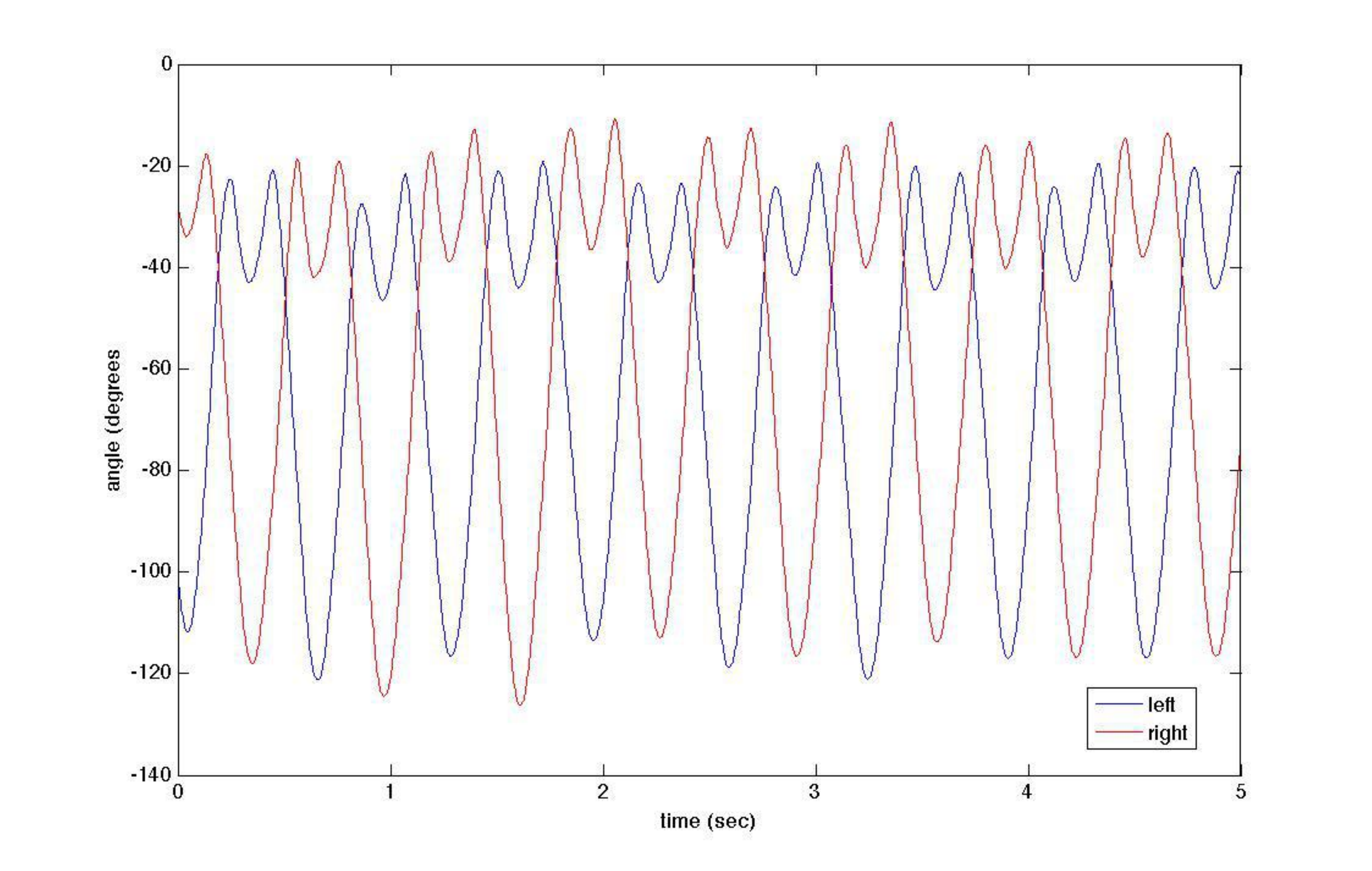}
     \includegraphics[width=0.8\textwidth]{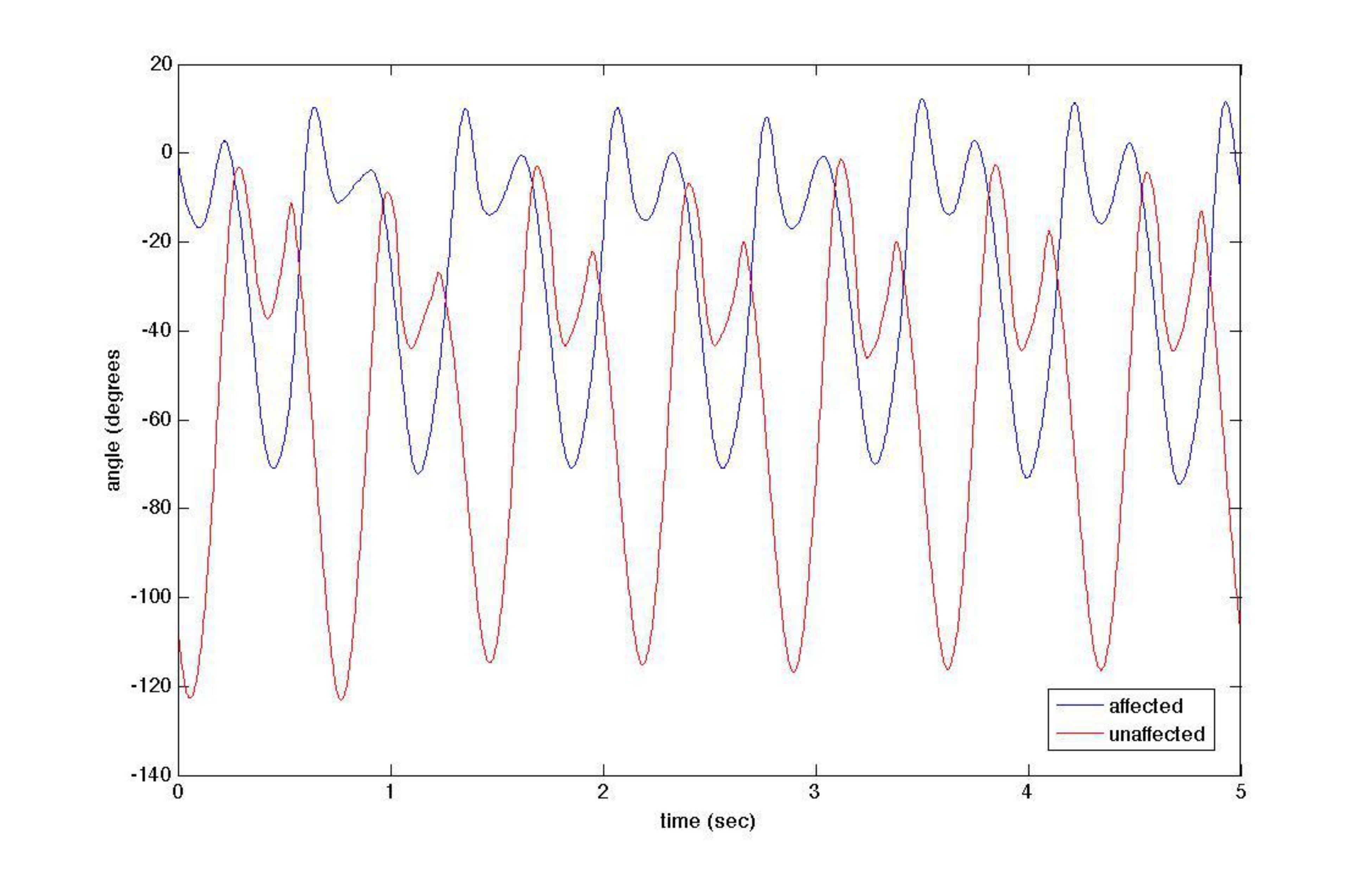}
     \caption{Sagittal-plane knee angles for (top) a non-amputee
       runner and (bottom) a runner with a unilateral transtibial
       amputation, both running at 4 m/s.  $0^{\circ}$ is full
       extension; negative angles correspond to flexion of the joint.}
 \label{fig:time-series-example}
 \end{figure}

Since the body is a coupled dynamical system, one should theoretically
be able to use delay-coordinate embedding to reconstruct its
$d$-dimensional dynamics from any single joint position (or angle).
Here, though, we wished to focus on smaller units of the body.
To this end, we used the medio-lateral ($x$), anterior-posterior ($y$)
and vertical ($z$) position coordinates of the sacrum marker to assess
the center-of-mass dynamics.  To explore the lower-limb dynamics, we
used the sagittal plane knee- and hip-joint angles.  Examples of these
data can be seen in Figure~\ref{fig:time-series-example}, which shows
traces of the knee-angle data from two of the runners in this study,
one NA and one \amp subject.
The temporal patterns in the left- and right-knee angles of the NA
runner are very similar, though they are of course roughly 180 degrees
out of phase.  There is an obvious difference, however, between the
knee angles of the affected and unaffected legs of the \amp subject.
All four traces---both knees of both runners---demonstrate largely,
but not completely, periodic motion.

\begin{figure}
  \centering
  \subfigure[NA mutual information]{\includegraphics[width=0.49\columnwidth]{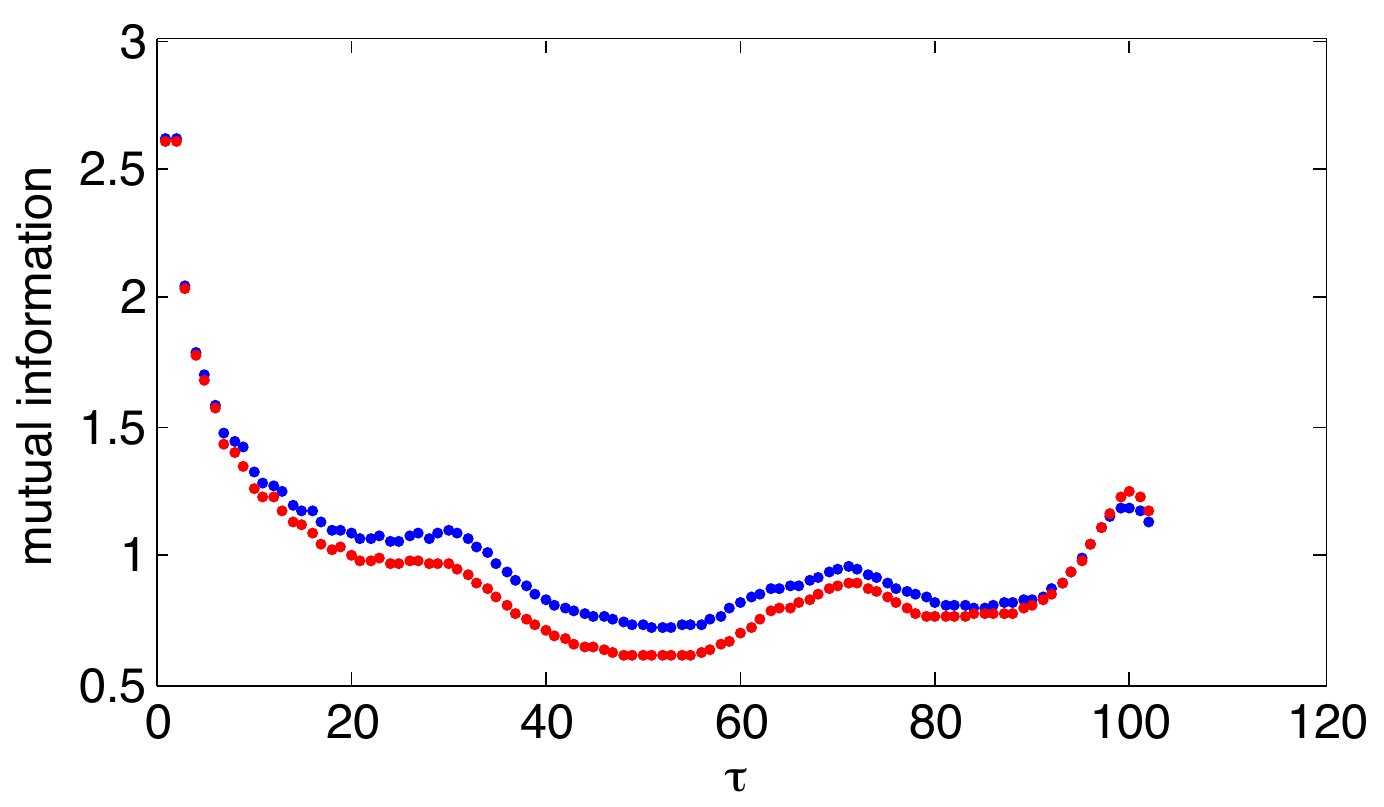}}
  \subfigure[NA false near neighbors]{\includegraphics[width=0.49\columnwidth]{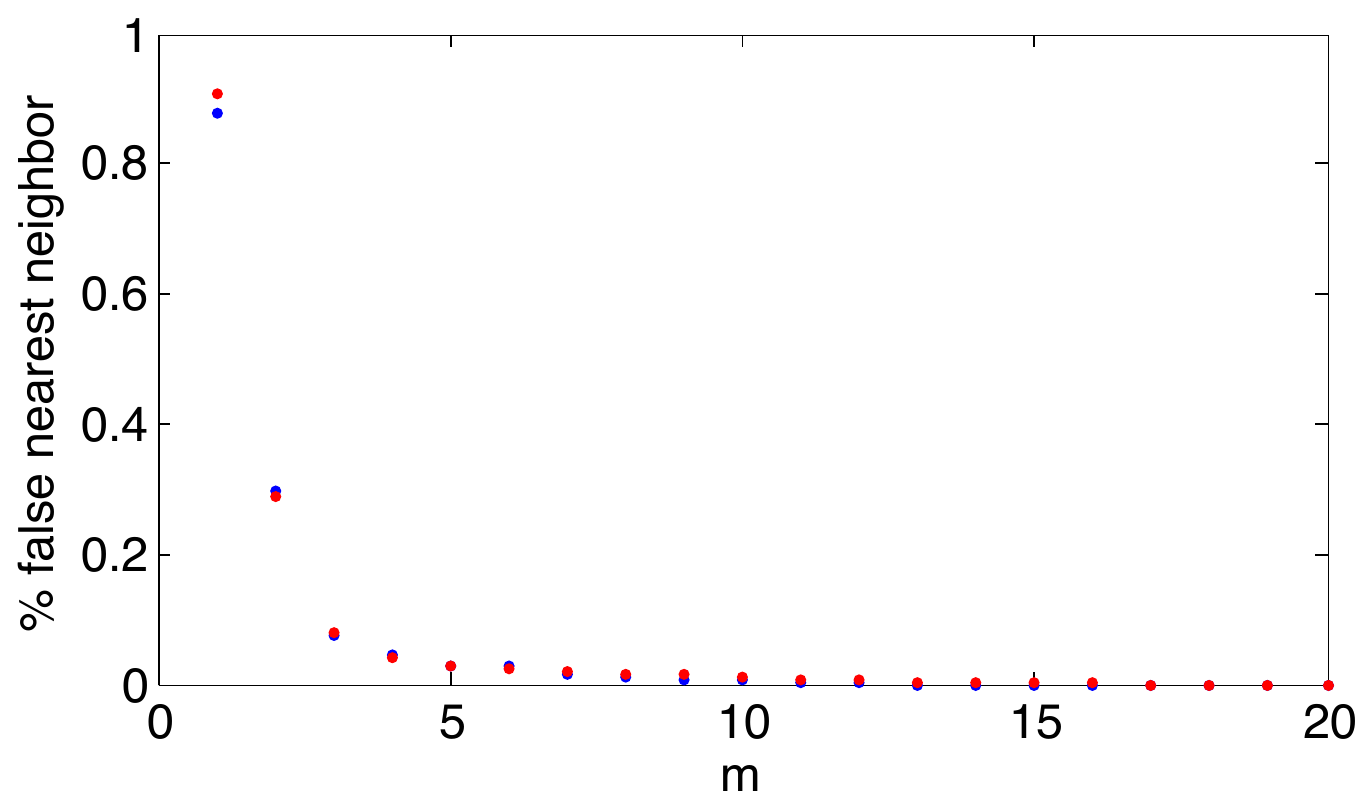}}
  \subfigure[\amp mutual information]{\includegraphics[width=0.49\columnwidth]{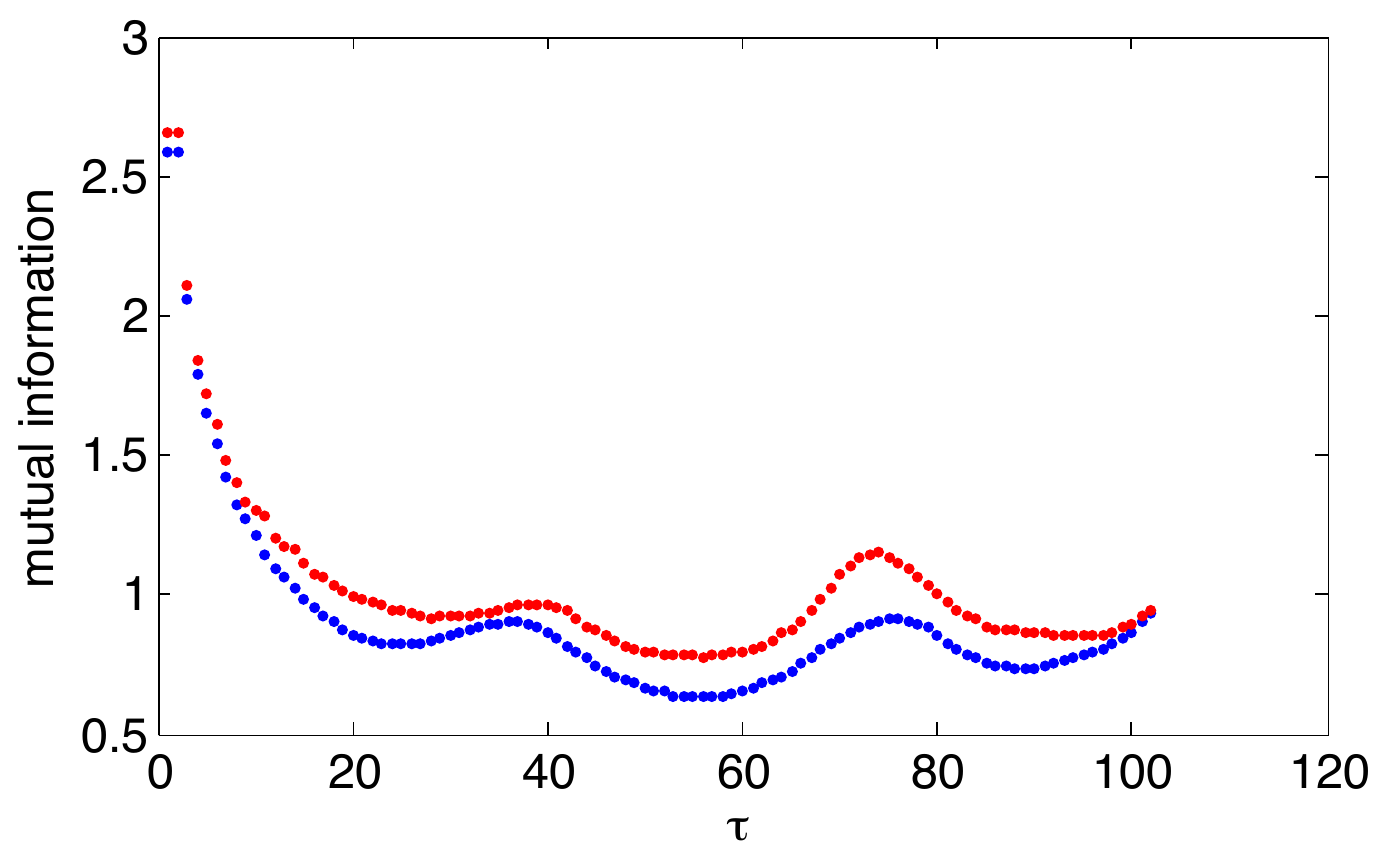}}
  \subfigure[\amp false near neighbors]{\includegraphics[width=0.50\columnwidth]{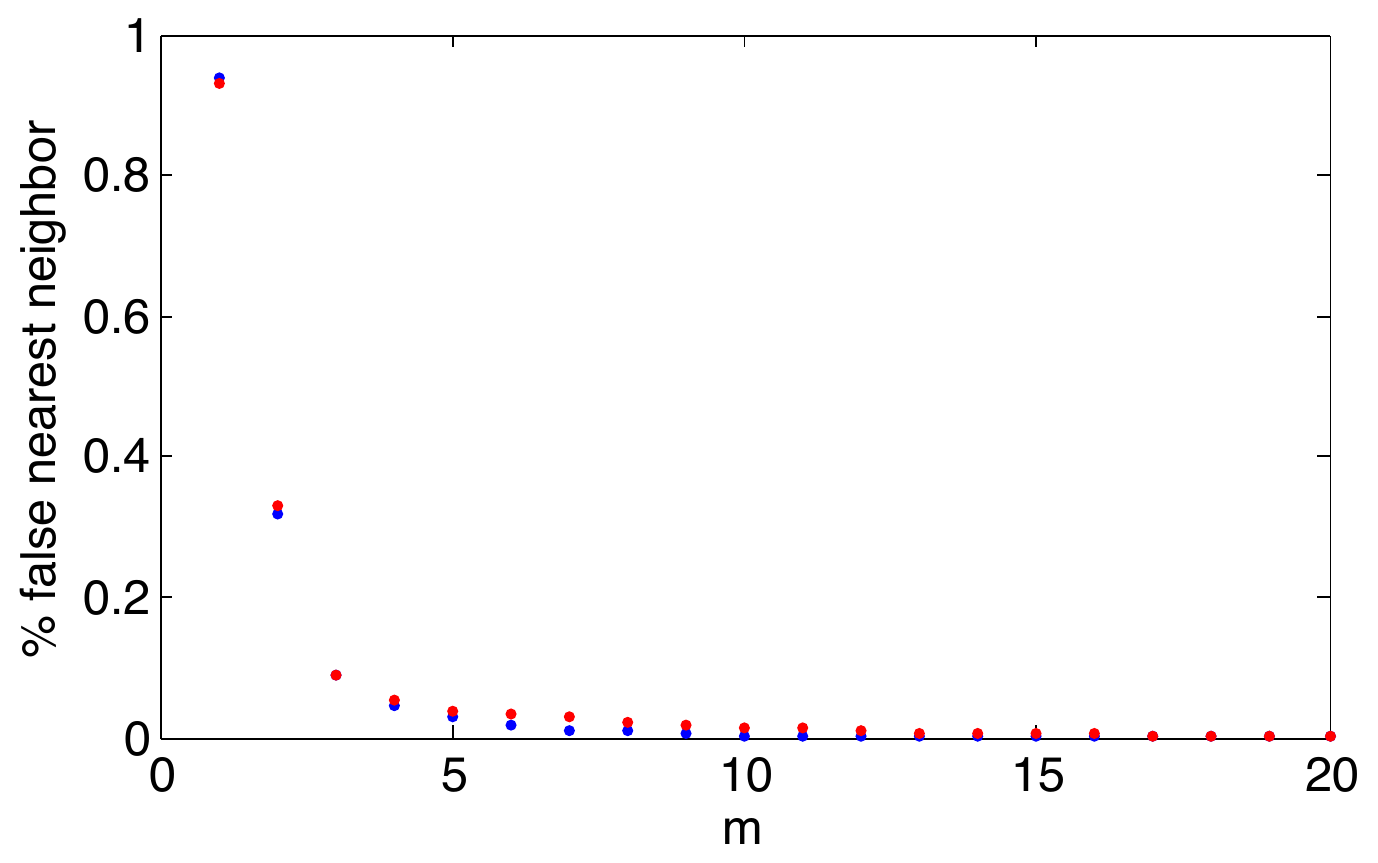}}
     \caption{Estimating embedding parameters for the data of
       Figure~\ref{fig:time-series-example}: mutual information as a
       function of the delay $\tau$, plotted in units of the sampling
       rate $\Delta t = 1/300^{th}$sec, and \% false near neighbors as
       a function of the dimension $m$.  The minima of the mutual
       information curves occur near $\tau=52 \Delta t$ (i.e., 173
       milliseconds) for both knees of the non-amputee (``NA'') runner
       and $\tau=55 \Delta t$ (i.e., 183 milliseconds) for both knees
       of the runner with a unilateral transtibial amputation (the
       ``WA'' subject).
%
%
       All four false near neighbor curves drop to 10\% at
       $m=3$. Color code as in the previous figure: blue and red
       correspond to the left and right leg, respectively, of the NA
       runner, and to the affected and unaffected leg, respectively,
       of the \amp subject. }
 \label{fig:embedding-parameters}
 \end{figure}

\begin{figure}
   \centering
  \subfigure[NA left knee ($\tau=21$; $m=3$)]
            {\includegraphics[width=0.49\columnwidth]{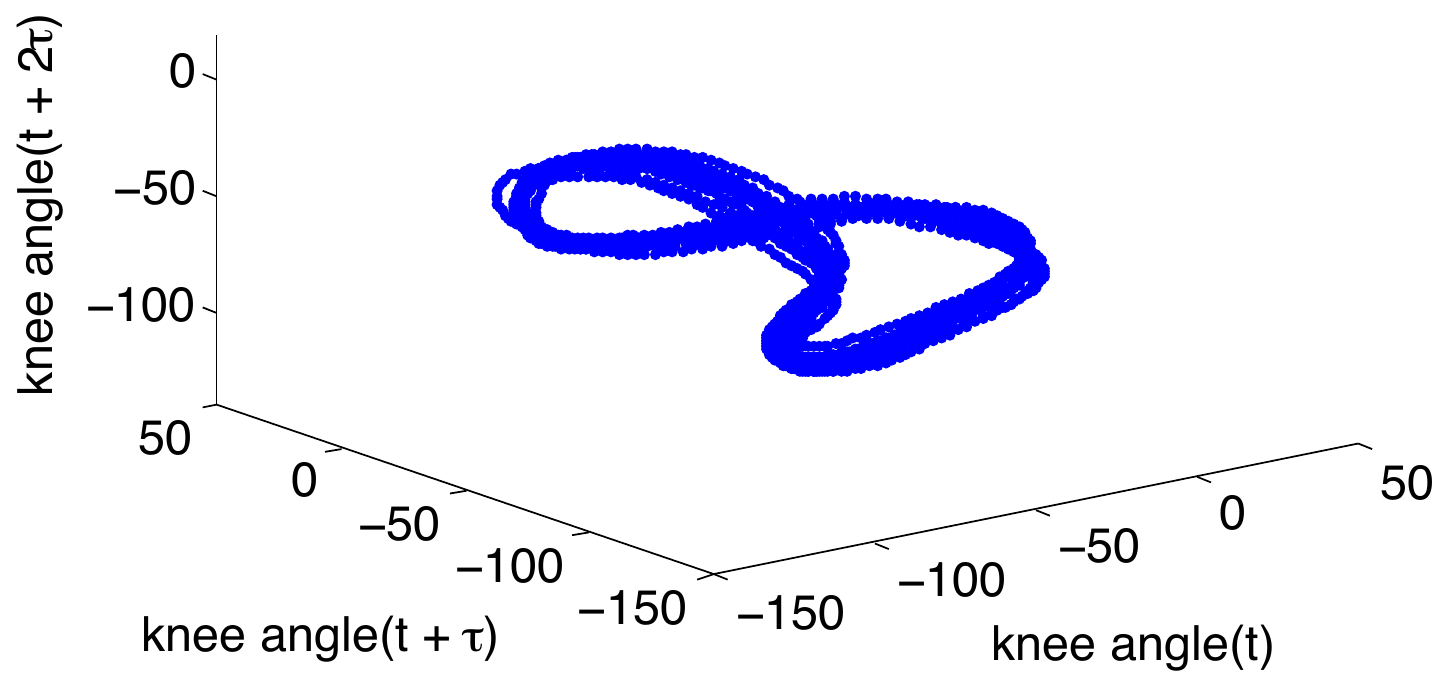}}
  \subfigure[NA right knee ($\tau=21$; $m=3$)]
            {\includegraphics[width=0.49\columnwidth]{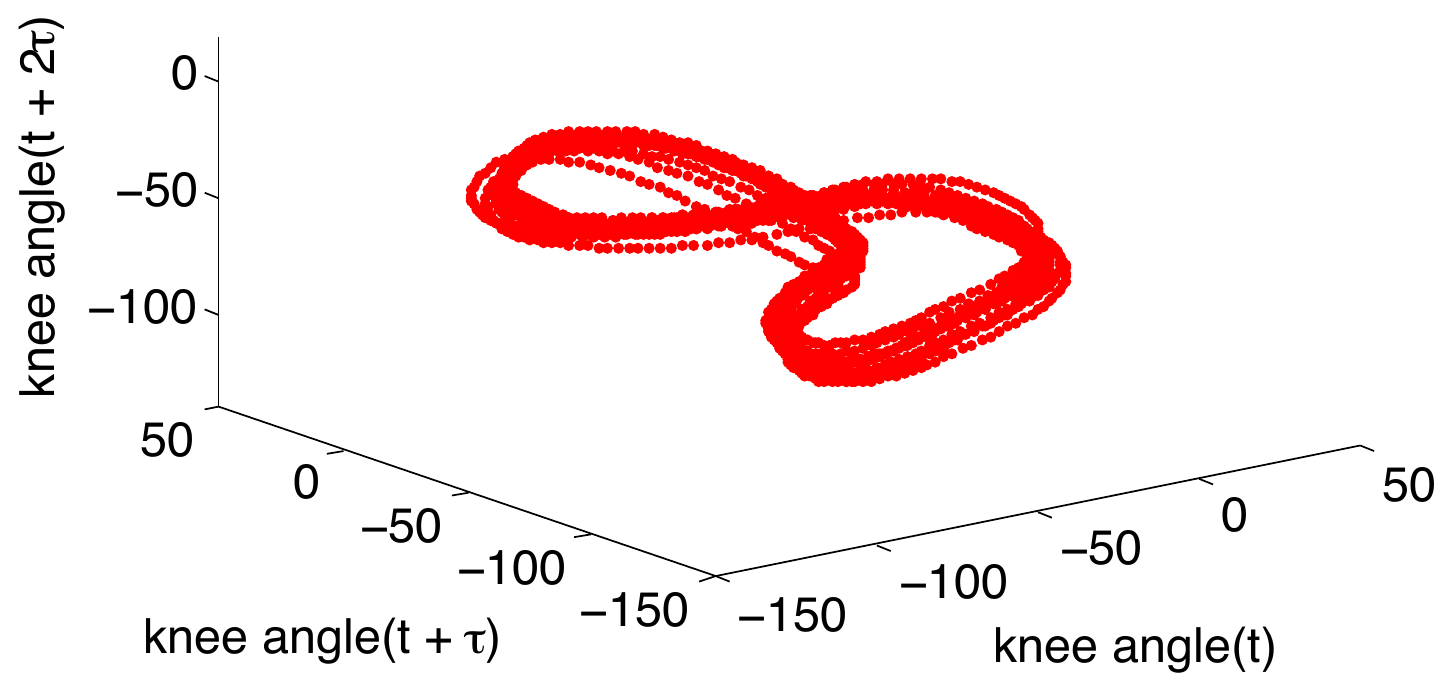}}
  \subfigure[\amp affected knee ($\tau=26$; $m=3$)]
            {\includegraphics[width=0.49\columnwidth]{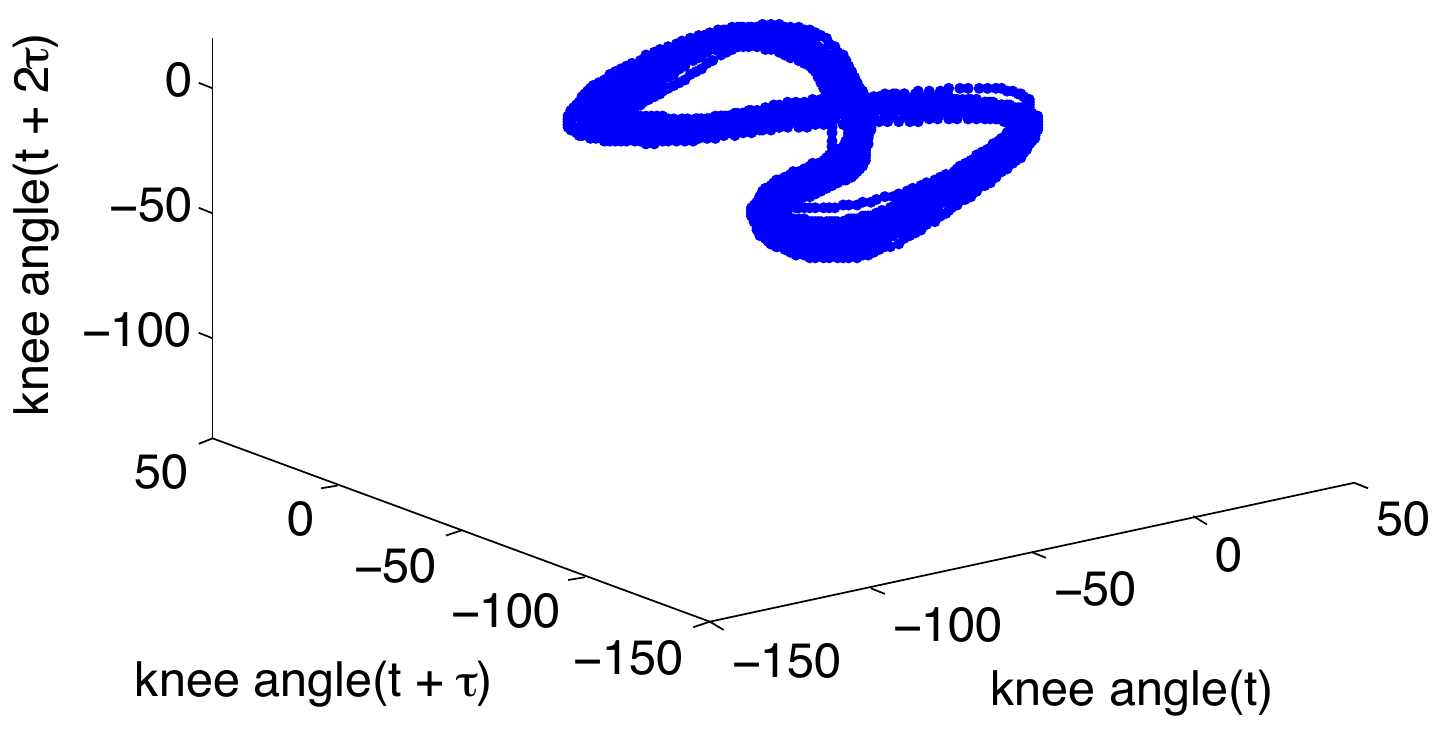}}
  \subfigure[\amp unaffected knee ($\tau=28$; $m=3$)]
            {\includegraphics[width=0.49\columnwidth]{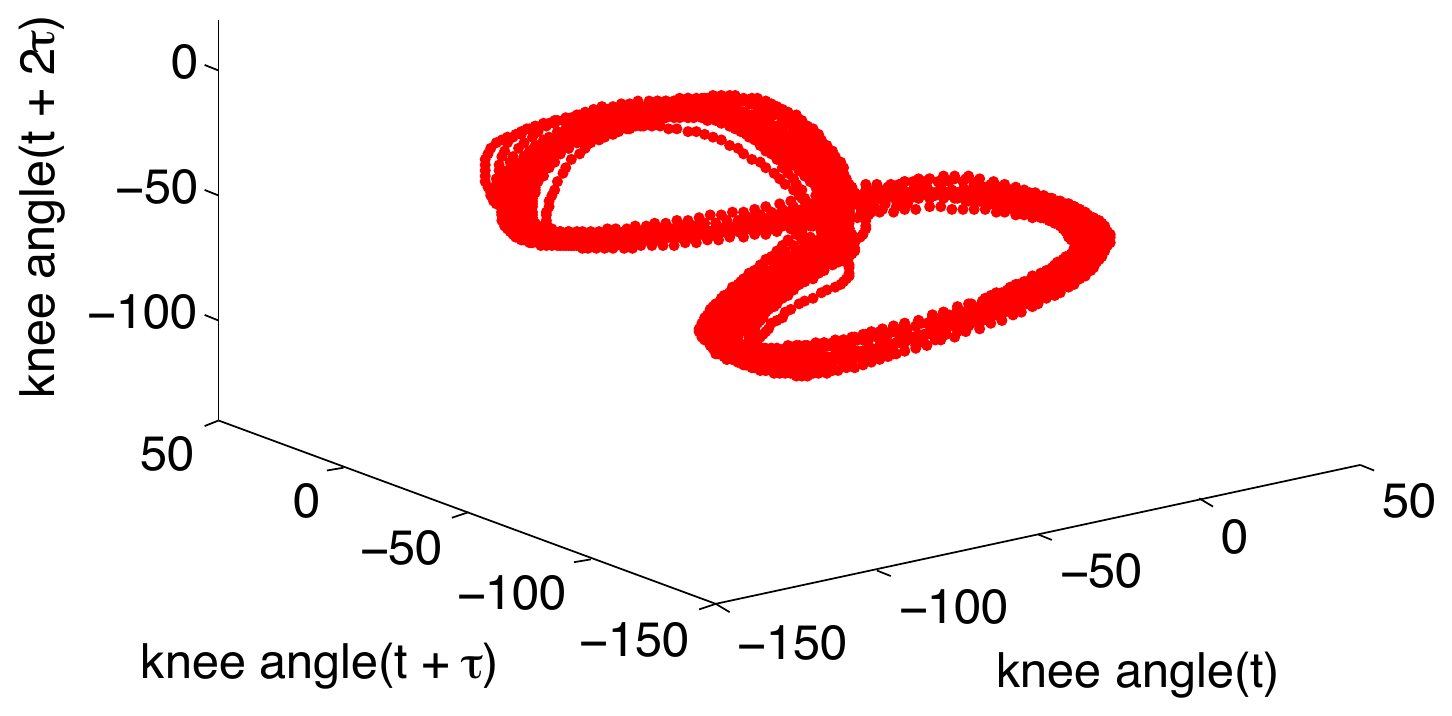}}
     \caption{Delay-coordinate embeddings of the traces in
       Figure~\ref{fig:time-series-example} with the $\tau$ and $m$
       values suggested by the curves in
       Figure~\ref{fig:embedding-parameters}. In these plots, time
       ($t$) is in units of $\Delta t$, the inverse of the 300 Hz
       sampling rate of the time series.}
 \label{fig:embedding-example}
 \end{figure}

To reconstruct the state-space dynamics from these time-series data,
we followed standard procedures regarding the choice of appropriate
values for the embedding parameters: the minimum of the
mutual-information curve \cite{fraser-swinney} as an estimate of the
delay $\tau$ and the false-near neighbors technique of \cite{KBA92},
with a threshold of 10\%, to estimate the embedding dimension $m$.
Figure~\ref{fig:embedding-parameters} shows the mutual information and
false-near neighbor curves for the time-series data of
Figure~\ref{fig:time-series-example}.
To perform these calculations, we used TISEAN's {\tt mutual} and {\tt
  false\_nearest} tools \cite{tisean-website}.  The corresponding
embeddings are shown in Figure~\ref{fig:embedding-example}.
For both legs of both subjects, $m = 3$ was sufficient to unfold the
dynamics and the first minima of the mutual information curves
occurred between $52\Delta t$ and $56\Delta t$, where the sampling
rate $\Delta t = 1/300$th of a second.  The embedded trajectories have
a characteristic figure-eight shape that reflects the general pattern
of running gait, but with visible stride-to-stride variations.

\begin{figure}
   \centering
  \subfigure[NA left knee]
            {\includegraphics[width=0.49\columnwidth]{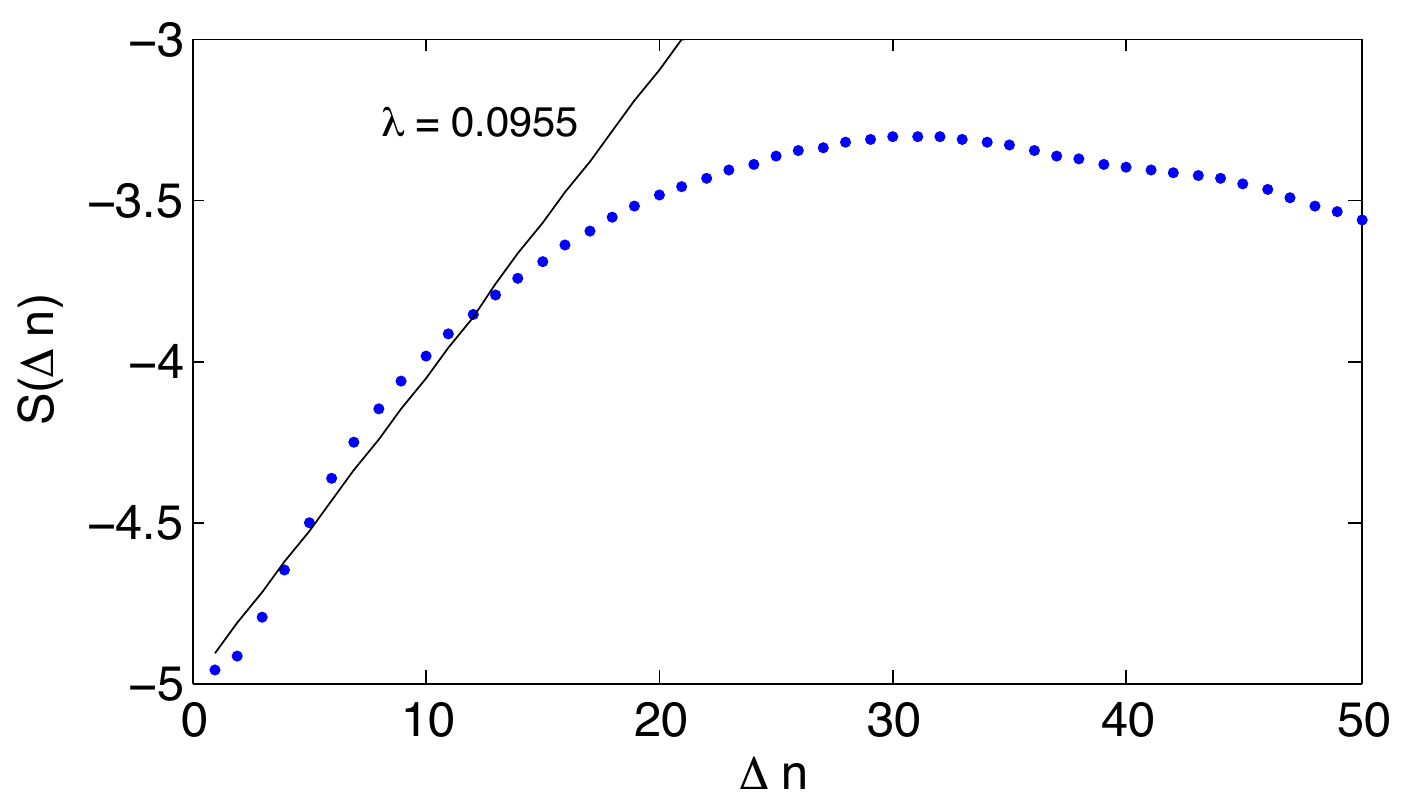}}
  \subfigure[NA right knee]
            {\includegraphics[width=0.49\columnwidth]{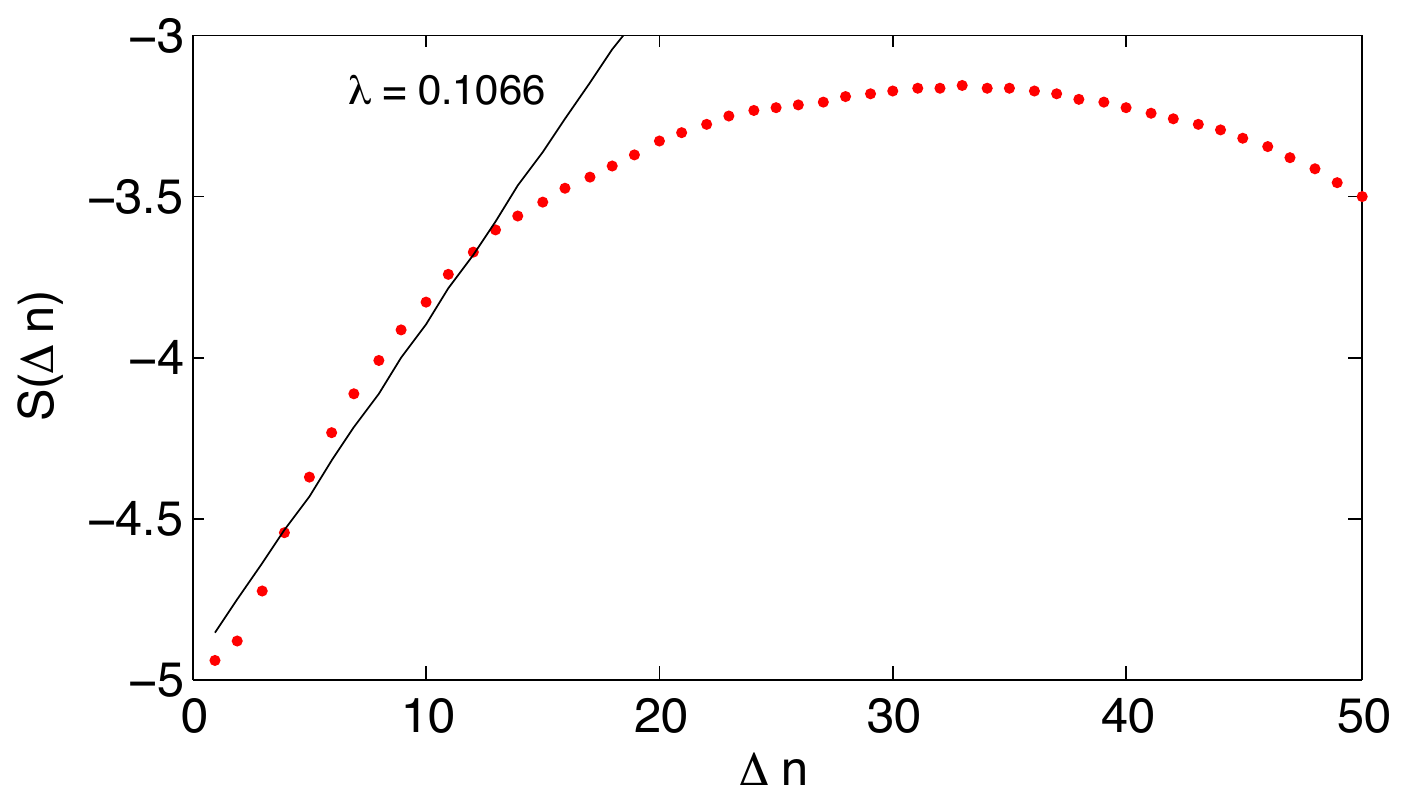}}
    \subfigure[\amp affected knee]
            {\includegraphics[width=0.49\columnwidth]{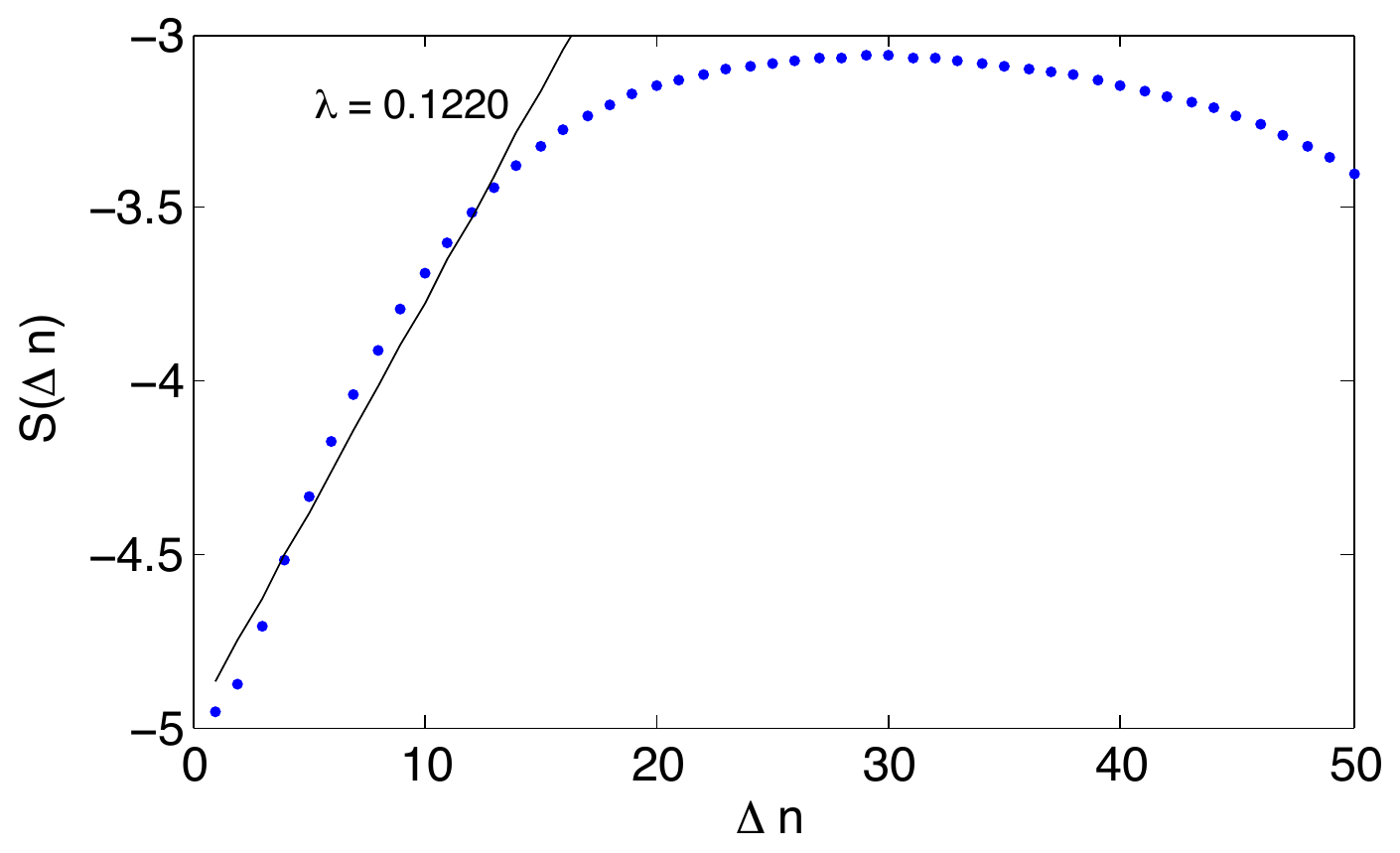}}
   \subfigure[\amp unaffected knee]
            {\includegraphics[width=0.49\columnwidth]{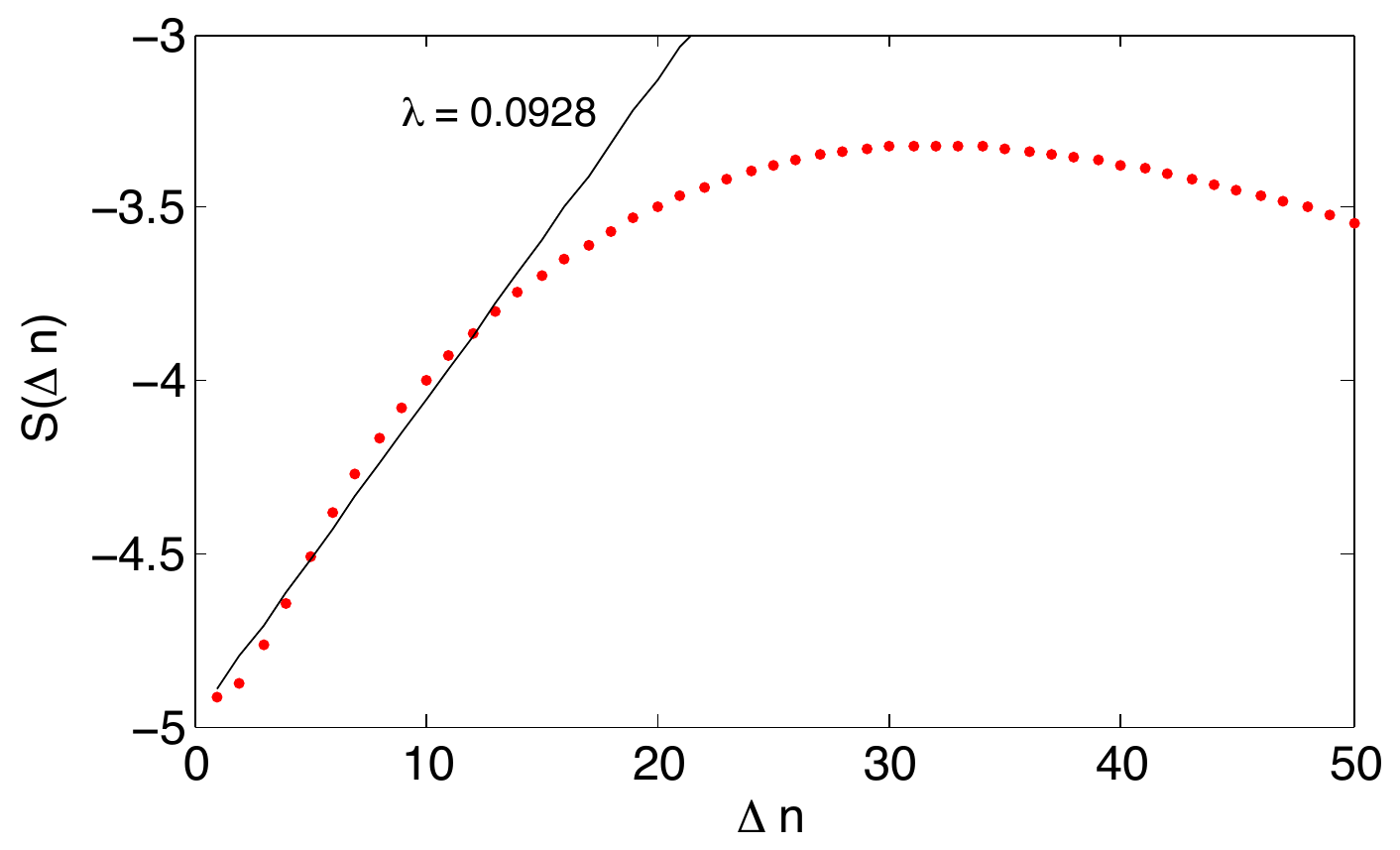}}
     \caption{Lyapunov exponent calculations for the embedded data of
       Figure~\ref{fig:embedding-parameters}.  The slopes of the
       scaling regions of these curves---fit by the superimposed lines
       in the plots---represent estimates of the maximal Lyapunov
       exponent $\lambda_1$ of the corresponding trajectories.
       $\Delta n$ is plotted in units of $\Delta t$, the inverse of
       the 300Hz sampling rate of the time series.}
 \label{fig:lambda-example}
 \end{figure}

To study these patterns and variations, we employed the algorithm of
Kantz \cite{kantz94}, as instantiated in TISEAN's {\tt lyap\_k} tool,
to estimate the maximal Lyapunov exponent $\lambda_1$ of the embedded
data.  First, we plotted the log of the expansion rate $S(\Delta n)$
versus time $\Delta n$ and verified that the curves were of the
approriate shape (i.e., a scaling region followed by a horizontal
asymptote, which should occur when the time horizon of the algorithm
is large enough to allow neighboring trajectories to stretch across
the diameter of the attractor).  We then fit a line to that scaling
region and determined its slope.  All $\lambda_1$ values in this paper
are scaled to the inverse of the sampling interval, $\Delta t =
1/300^{th}$ second; to convert these $\lambda_1$ values to inverse
seconds, one multiplies these values by the sampling frequency (300).
Figure~\ref{fig:lambda-example} shows the $ \log S(\Delta n)$ versus
$\Delta n$ curves for the time-series data in
Figure~\ref{fig:time-series-example}, embedded using the $\tau$ and
$m$ values suggested by the curves in
Figure~\ref{fig:embedding-parameters}.
These results indicated that the dynamics of both knees of each of
these two runners was sensitively dependent on initial conditions,
with $\lambda_1$ ranging from 0.0084--0.0154 per $\Delta t$, which
translates to 2.52--4.62 in units of inverse seconds.  In both
subjects, the $\lambda_1$ values differed between the two legs, but
the difference was more pronounced for the \amp subject.  This pattern
is discussed at more length in the following section.

All of the nonlinear time-series analysis algorithms mentioned in this
section are known to be sensitive to data and parameter effects
\cite{kantz97}.  These systematic uncertainties preclude the use of
traditional statistics to assess or compare their results, but there
are other ways to do ``due diligence.''  We validated all $\lambda_1$
calculations by repeating them for a range of values of the dimension
$m$ and the critical scale parameter ($\epsilon$) in the {\tt lyap\_k}
algorithm\footnote{This parameter specifies the size of the
  neighborhood whose points are tracked for the calculation of the
  spreading factor $S$.  Too-small values of $\epsilon$ cause
  numerical problems because the neighborhood contains only a few
  points; too-large values cause the calculation to sample the
  dynamics too broadly.}  and discarding any that produced
inconsistent results (i.e., large variation with $m$ and/or
$\epsilon$).  We also discarded all results from $S(\Delta n)$ versus
$\Delta n$ curves that did not have a clear scaling region.  We
repeated all of these calculations on seven traces (right and left
knee- and hip-joint angles, plus the $x$, $y$, and $z$ positions of
the sacrum) for each of the 17 subjects.  The following section
summarizes the results and examines the differences and similarities
between and across subjects, groups, speeds, and legs.

\section{Results}
\label{sec:results}

\subsection{Knee Dynamics}

Our analysis of the embedded knee-joint dynamics supports our first
hypothesis: stability decreases (viz., increasing $\lambda_1$) at
faster speeds for all runners, with and without amputations.  The
average $\lambda_1$ of the right and left knee angles of the NA
runners increased from 0.095 and 0.101 at 3 m/s to 0.137 and 0.136 at
9 m/s, respectively.  The overall speed-stability trends---higher
$\lambda_1$ at faster running speeds---were similar for the \amp
subjects, further supporting our first hypothesis.  The average
$\lambda_1$ of \amp subjects were 0.103 and 0.090 for affected and
unaffected legs, respectively, at 3 m/s; at 9 m/s, the corresponding
values were 0.138 and 0.124.  See Table~\ref{tab:knee-lambda-vs-speed}
for $\lambda_1$ values for both groups across all running speeds.
\begin{table}
   \centering
\begin{tabular}{|c|c|c|c|}
\hline
\multicolumn{4}{|l|}{{\bf NA subjects}} \\
\hline
Speed & Sample & Left & Right \\
(m/s) & Size & $\lambda_1$ & $\lambda_1$ \\
\hline
3 & 9 & 0.101 $\pm$ 0.016 & 0.095 $\pm$ 0.017 \\
4 & 11 & 0.096 $\pm$ 0.017 & 0.096 $\pm$ 0.016 \\
5 & 10 & 0.100 $\pm$ 0.009 & 0.104 $\pm$ 0.009 \\
6 & 10 & 0.110 $\pm$ 0.020 & 0.113 $\pm$ 0.020 \\
7 & 10 & 0.113 $\pm$ 0.020 & 0.115 $\pm$ 0.016 \\
8 & 8 & 0.117 $\pm$ 0.012 & 0.121 $\pm$ 0.013 \\
9 & 8 & 0.136 $\pm$ 0.016 & 0.137 $\pm$ 0.007 \\
\hline
\end{tabular}
\begin{tabular}{|c|c|c|c|}
\hline
\multicolumn{4}{|l|}{{\bf \amp subjects}} \\
\hline
Speed & Sample & Unaffected & Affected \\
(m/s) & Size & $\lambda_1$ & $\lambda_1$ \\
\hline
3 & 4 & 0.098 $\pm$ 0.018 & 0.103 $\pm$ 0.009 \\
4 & 5 & 0.103 $\pm$ 0.012 & 0.107 $\pm$ 0.009 \\
5 & 5 & 0.102 $\pm$ 0.102 & 0.124 $\pm$ 0.019 \\
6 & 6 & 0.104 $\pm$ 0.104 & 0.118 $\pm$ 0.017 \\
7 & 5 & 0.113 $\pm$ 0.113 & 0.133 $\pm$ 0.017 \\
8 & 4 & 0.119 $\pm$ 0.119 & 0.137 $\pm$ 0.024 \\
9 & 4 & 0.124 $\pm$ 0.124 & 0.138 $\pm$ 0.037 \\
\hline
\end{tabular}
     \caption{$\lambda_1$ values for the embedded knee-joint dynamics
       of non-amputees and subjects with amputations.  The values
       reported are averages across all traces in the corresponding
       class at that speed (e.g., the average of the right-knee
       $\lambda_1$ values of all NA runners at 3 m/s was 0.095 in
       units of inverse $\Delta t$, the 300Hz sampling interval of the
       data, with a standard deviation of 0.017).}
 \label{tab:knee-lambda-vs-speed}
\end{table}

The symmetry of the system---the subject of the second hypothesis---is
reflected in the similarities and differences between the numbers in
Table~\ref{tab:knee-lambda-vs-speed}.  As one would expect, the left
and right knee dynamics were quite similar in the NA subjects, who
have two intact biological legs.  Not surprisingly, there were
differences between legs in the \amp subjects.  This result is
consistent with our second hypothesis regarding inter-leg asymmetry in
this group.  

\subsection{Hip Dynamics}

The $\lambda_1$ values for the reconstructed hip-joint dynamics were
also consistent with our first two hypotheses.  The average
$\lambda_1$ of NA runners were 0.098 and 0.103 at 3 m/s for the right
and left legs, respectively; these values increased to 0.119 and 0.116
at 9 m/s (Table~\ref{tab:hip-lambda-vs-speed}).
\begin{table}
   \centering
\begin{tabular}{|c|c|c|c|}
\hline
\multicolumn{4}{|l|}{{\bf NA subjects}} \\
\hline
Speed & Sample & Left & Right \\
(m/s) & Size & $\lambda_1$ & $\lambda_1$ \\
\hline
3 & 9 & 0.103 $\pm$ 0.028 & 0.098 $\pm$ 0.013 \\
4 & 11 & 0.098 $\pm$ 0.023 & 0.095 $\pm$ 0.015 \\
5 & 10 & 0.097 $\pm$ 0.020 & 0.108 $\pm$ 0.016 \\
6 & 10 & 0.107 $\pm$ 0.015 & 0.100 $\pm$ 0.023 \\
7 & 10 & 0.103 $\pm$ 0.017 & 0.106 $\pm$ 0.018 \\
8 & 8 & 0.112$\pm$ 0.008 & 0.110 $\pm$ 0.022 \\
9 & 8 & 0.116 $\pm$ 0.010 & 0.119 $\pm$ 0.014 \\
\hline
\end{tabular}
\begin{tabular}{|c|c|c|c|}
\hline
\multicolumn{4}{|l|}{{\bf \amp subjects}} \\
\hline
Speed & Sample & Unaffected & Affected \\
(m/s) & Size & $\lambda_1$ & $\lambda_1$ \\
\hline
3 & 4 & 0.075 $\pm$ 0.020 & 0.113 $\pm$ 0.017 \\
4 & 5 & 0.097 $\pm$ 0.014 & 0.116 $\pm$ 0.017 \\
5 & 5 & 0.095 $\pm$ 0.010 & 0.109 $\pm$ 0.017 \\
6 & 6 & 0.098 $\pm$ 0.009 & 0.117 $\pm$ 0.021 \\
7 & 5 & 0.104 $\pm$ 0.012 & 0.115 $\pm$ 0.014 \\
8 & 4 & 0.111 $\pm$ 0.011 & 0.119 $\pm$ 0.014 \\
9 & 4 & 0.122 $\pm$ 0.004 & 0.131 $\pm$ 0.005 \\
\hline
\end{tabular}
     \caption{$\lambda_1$ values for the embedded hip-joint dynamics
       of non-amputees and subjects with amputations.  The values
       reported are averages across all traces in the corresponding
       class at that speed (e.g., the average of the right-hip
       $\lambda_1$ values of all NA runners at 3 m/s was 0.098 in
       units of inverse $\Delta t$, the 300Hz sampling interval of the
       data, with a standard deviation of 0.013).}
 \label{tab:hip-lambda-vs-speed}
\end{table}
Again, the values for the right and left legs were similar for the NA
subjects, reflecting the symmetry in their running gait.  As in the
case of the knee-angle results in the previous section, the embedded
hip-joint data provide some indications of asymmetry in the dynamics
between unaffected and affected legs of the \amp subjects, again
supporting our second hypothesis.  The average $\lambda_1$ of \amp
subjects were 0.075 and 0.113 at 3 m/s for the affected and unaffected
legs, respectively; these values increased to 0.122 and 0.131 at 9
m/s.  This convergence of $\lambda_1$ values at higher running
speeds---a reduction in the asymmetry in the dynamics---might indicate
that while there could be many different mechanical choices to run
slowly (i.e., fewer constraints), there may only be one effective way
to run at faster speeds.  The average $\lambda_1$ increased at faster
running speeds for both legs in both groups, again supporting our
first hypothesis.  Since the knee and hip angles are measurements of
the same dynamical system---essentially, different measurement
functions $h$ applied to the same underlying dynamics $M$---these
corroborations are not surprising.  We discuss this at more length in
Section~\ref{sec:discussion}.

\subsection{Center-of-Mass Dynamics}

As discussed on page~\pageref{page:COM-sacrum}, we used the sacrum
marker at the base of the spine as a proxy for the center-of-mass
position.  See Table~\ref{tab:sacrum-lambda-vs-speed} for $\lambda_1$
values for the dynamics reconstructed from the medio-lateral and
vertical positions of this marker.  (The anterior-posterior position
of the sacrum during treadmill running reflects more about the
subjects' ability to match treadmill speed than anything else, and
hence was not included in these analyses.)
%
%
\begin{table}
   \centering
\begin{tabular}{|c|c|c|c|}
\hline
\multicolumn{4}{|l|}{{\bf NA subjects}} \\
\hline
Speed & Sample & Medio-Lateral & Vertical \\
(m/s) & Size & $\lambda_1$ & $\lambda_1$ \\
\hline
3 & 9 & 0.016 $\pm$ 0.006 & 0.130 $\pm$ 0.016 \\
4 & 11 & 0.031 $\pm$ 0.010 & 0.121 $\pm$ 0.021 \\
5 & 10 & 0.034 $\pm$ 0.018 & 0.130 $\pm$ 0.016 \\
6 & 10 & 0.039 $\pm$ 0.015 & 0.142 $\pm$ 0.035 \\
7 & 10 & 0.095 $\pm$ 0.040 & 0.127 $\pm$ 0.062 \\
8 & 8 & 0.100 $\pm$ 0.041  & 0.104 $\pm$ 0.066 \\
9 & 8 & 0.128 $\pm$ 0.015  & 0.128 $\pm$ 0.044 \\
\hline
\end{tabular}
\begin{tabular}{|c|c|c|c|}
\hline
\multicolumn{4}{|l|}{{\bf \amp subjects}} \\
\hline
Speed & Sample & Medio-Lateral & Vertical \\
(m/s) & Size & $\lambda_1$ & $\lambda_1$ \\
\hline
3 & 4 & 0.015 $\pm$ 0.011 & 0.100 $\pm$ 0.031 \\
4 & 5 & 0.046 $\pm$ 0.040 & 0.123 $\pm$ 0.033 \\
5 & 5 & 0.051 $\pm$ 0.039 & 0.114 $\pm$ 0.038 \\
6 & 6 & 0.073 $\pm$ 0.041 & 0.119 $\pm$ 0.056 \\
7 & 5 & 0.074 $\pm$ 0.043 & 0.087 $\pm$ 0.053 \\
8 & 4 & 0.093 $\pm$ 0.056 & 0.084 $\pm$ 0.051 \\
9 & 4 & 0.084 $\pm$ 0.062 & 0.052 $\pm$ 0.012 \\
\hline
\end{tabular}
     \caption{$\lambda_1$ values for the embedded sacrum-position
       dynamics of non-amputees and subjects with amputations.  The
       $\lambda_1$ values reported in the two right-hand columns are
       averages across all traces in the corresponding class at that
       speed (e.g., the average of the medio-lateral $\lambda_1$
       values of all NA runners at 3 m/s was 0.016 in units of inverse
       $\Delta t$, the 300Hz sampling interval of the data, with a
       standard deviation of 0.006).}
 \label{tab:sacrum-lambda-vs-speed}
\end{table}
The $\lambda_1$ of the medio-lateral dynamics of the sacrum marker
generally increased with running speed, which is in accordance with
our first hypothesis.
However, the dynamics reconstructed from time-series data of the {\sl
  vertical} position of the sacrum exhibited a different pattern.  In
NA runners, the $\lambda_1$ of these embedded dynamics did not show
any clear pattern with increasing speed; in \amp runners, $\lambda_1$
{\sl decreased} with speed.  
%
%
These patterns are significantly different from those in the hip- and
knee-joint dynamics.  Since all of these data are simultaneous
measurements of different macroscopic variables in the same dynamical
system, this discrepancy between joint dynamics and center-of-mass
dynamics is a puzzling finding from a dynamical-systems standpoint;
see the following section for more discussion of this issue.

The sacrum position data also had interesting implications regarding
our third hypothesis (that the center-of-mass dynamics of \amp runners
will be less stable than in NA runners\footnote{The second hypothesis
  is not at issue in this section, since the sacrum position data do
  not effectively isolate the dynamics of the individual lower
  limbs.}).  The answer appears not to be so simple.  Across all
speeds, $\lambda_1$ was smaller in the vertical dynamics for \amp
runners---i.e., those dynamics were more stable.  In the medio-lateral
direction, \amp runners were more stable than NA runners at slower
speeds, but at faster speeds, NA runners were more stable.


\section{Discussion}
\label{sec:discussion}

A nonlinear time-series analysis of knee, hip, and sacrum dynamics of
runners with and without a unilateral transtibial amputation confirmed
our hypothesis that $\lambda_1$ generally increases with running
speed, with one exception: the vertical position of the sacrum, where
the $\lambda_1$ of the embedded data {\sl decreased} with running
speed for \amp subjects and remained roughly the same for NA subjects.
We have two conjectures about this result:
%
%
\begin{itemize}
\item It may be the case that runners exert increased control of the
core to balance decreased stability elsewhere, and this effect may be
more pronounced in \amp runners.
\item Because running involves very little side-to-side motion, the
  associated vertical movement of the sacrum may dominate the
  dynamics.  That is, there could be a strongly stable limit cycle in
  the vertical sacrum dynamics due to the mechanical energy
  storage/return mechanism that governs vertical interactions with the
  ground.
\end{itemize}

Our second hypothesis concerned symmetry in the embedded lower-limb
dynamics for \amp subjects.  Our analysis indicated that the
$\lambda_1$ of the embedded time-series data from the affected leg was
indeed higher than for the unaffected leg---except for knee data at
low speeds.  

Our third hypothesis---that the $\lambda_1$ of the center-of-mass
dynamics of \amp runners would be higher than for NA runners---was
{\sl not} verified by this analysis, except for a few midrange speeds
in the medio-lateral sacrum position data.  This may be due to the
effects discussed in the first paragraph of this section.  It is also
important to note that we observed small amounts of nonstationarity in
the medio-lateral data due to subtle changes of the subject’s position
on the treadmill.  Although others have minimized nonstationarities in
the signal using divided difference methods prior to computing
Lyapunov exponents \cite{dingwell06}, we avoided that approach because
those methods amplify any inherent noise in the signal.  In addition,
we believe that these slight nonstationarities represents behavior
that is dynamically meaningful, as opposed to the kind of unavoidable
drift that occurs in a measurement sensor.  In our view, slight
changes in the subject's position from step to step may represent
responses to local disturbances during running, thus providing
additional insight into dynamic stability.

Readers from the biomechanics community will have noted that we did
not do any of the traditional statistics analyses on these
results---e.g., fitting a regression line to the data in the tables
and giving an $R^2$ value to quantify our certainty about whether or
not those data validate a particular hypothesis.  Numerical algorithms
that extract important properties of complicated nonlinear dynamical
systems are based on approximations of the associated theory.  They
involve a number of parameters that strongly affect the results, and
they are notoriously sensitive to noise, data length, and other
sampling effects.  A systematic exploration of these effects is
mandatory if one is to believe the results: minimally, a comparison of
the results of different algorithms and a careful exploration of the
parameter space of each one.  (We performed all of these kinds of
checks on our results.)  And since algorithms like {\tt lyap\_k}
inject systematic biases in the results, the underlying assumptions of
traditional statistics---e.g., normally distributed errors---do not
hold\footnote{Indeed, Kantz \& Schreiber \cite{kantz97} quote Salman
  Rushdie to make this point.}.

At this point, we are unaware of any other studies that have
quantified the nonlinear dynamics of time-series data for individuals
with unilateral amputations running across a range of speeds, as
described here.  Our findings on this unique population of runners,
then, are difficult to compare directly to other work.  Enoka {\sl et
  al.} \cite{enoka} were the first to provide important insights into
the asymmetries that exist between the biological and prosthetic leg
in individuals with unilateral amputations.  As was normal in that
era, the runners with amputations used inelastic prostheses designed
for walking, not running.  Yet, they were able to run at speeds
ranging from 2.7 m/s--8.2 m/s and exhibited notable kinematic
intra-limb asymmetries, e.g. significant reductions in the joint angle
range of motion of the prosthetic leg compared to the biological leg.
The leg prostheses used by runners in our study were designed to mimic
the spring-like mechanical behavior of biological legs more closely.
Even so, we observed slight asymmetries in the stability of the hip
and knee dynamics.  We also observed slight asymmetries in the
stability of the hip and knee dynamics, indicating that the use of
running-specific prostheses do not yet exactly replicate the
biomechanical function of biological legs.


Readers from the nonlinear dynamics community will have noted the
differences between the $\lambda_1$ values.  This bears some
explanation since the different time-series data sets studied here are
simultaneous samples of the same nonlinear dynamical system.
Theoretically, the $\lambda_1$ values of the dynamics reconstructed
from these different time-series datasets should be the same.  This
holds if the sensors that measure those different angles effect
smooth, generic functions of at least one state variable of that
system, and as long as the dynamics themselves are smooth.  In
practice, the length of the datasets plays a role as well.  If the
dynamical coupling between parts of the body is weak, that coupling
will not manifest during a short time series and thus the
``invariants'' of the reconstructed dynamics will not be the same from
joint to joint.  Since we were interested in the dynamics of gaits
that could not be sustained indefinitely (viz., running at top speed),
gathering longer time series was not an option.  The $\lambda_1$
values reported here, then, are really more like local $\lambda$s
\cite{local-lambdas}, also known as finite-time Lyapunov exponents,
and they should not be expected to be identical across the entire
body.  There may be other effects at work here: the sharp
ground-contact forces of running may disrupt the smoothness of the
dynamics, and the movement of the residual limb within the prosthetic
socket (``pistoning'') may add dynamics.


This study raises a variety of interesting questions regarding
stability, symmetry, and the effects of a running-specific prosthesis.
Here, we follow the practice of defining dynamic stability as the
resistance to a perturbation \cite{dingwell11}.  It is not clear,
however, whether the body reacts differently to endogenous
vs. exogenous perturbations.  The analysis presented here studies the
stability of running by analyzing the rate of divergence of nearby
trajectories in the reconstructed state space.  This provides some
indication of how the system responds to local perturbations, but it
does not distinguish between internal and external perturbations.  If
the system is autonomous, this distinction is irrelevant.  However, if
the dynamics are nonstationary---if the forward evolution from a given
point in state space depends on how (or when) one got to that
point---this distinction may be very important.  One could explore
this by delivering controlled perturbations to the subject on the
treadmill and studying the resulting dynamics.  These experiments
would be challenging.  There are a number of technical issues
surrounding sampling movement trajectories following a perturbation,
including the short time scales over which the body's internal
controller reacts and the potential hystereses and nonstationarities
in that controller: e.g., a shift from feedback to a feedforward
control strategy.  For instance, the body could learn, over time, to
prepare an appropriate response at the expected time of a
perturbation.  Experiments that could elucidate these effects, while
challenging, have the potential to reveal general strategies of how
the body's internal controller deals with external pertubations and
whether these responses can be captured by nonlinear time-series
analysis.
%

With regards to dynamic symmetry, the anthropomorphic differences
between the affected and unaffected legs of runners with a unilateral
transtibial amputation are accompanied by slight asymmetries in
stepping kinematics of running and sprinting \cite{grabowski10}.
Interestingly, adding mass ($\sim$300 g) to the running-specific
prosthesis helps to improve kinematic symmetry \cite{grabowski10}.
Similarly, anthromorphic and mass differences between the unaffected
and affected leg may create stability asymmetries in the dynamics of
runners with a unilateral amputation.  Our analysis suggests that the
ability to respond to small perturbations during running may be
compromised in the affected leg as compared to the unaffected leg.  It
is important to note that the running-specific prosthesis plus the
socket together weigh $\sim$2.3 kg while the biological leg (foot and
shank) weighs $\sim$3.6 kg.  The question remains as to whether adding
mass to the running-specific prosthesis, as explored in
\cite{grabowski10}, would improve the dynamic symmetry between the
unaffected and affected leg in runners with a unilateral amputation.

\bibliography{bib}

\end{document}